# Formalizing Oracle Trust Models for Blockchain-Based Business Applications. An Example from The Supply Chain Sector.

Giulio Caldarelli

University of Verona, Department of Business Administration

giulio.caldarelli@univr.it

ORCID: https://orcid.org/0000-0002-8922-7871

## Abstract

Blockchain technology truly opened the gate to a wave of unparalleled innovations; however, despite the rapidly growing load of hype, the integration into the business, apart from a few applications, seems to be coming at a slower rate. One reason for that delay may be the need in the real-world applications for the so-called "trust model." Trust models are rarely mentioned in blockchain application proposals despite their importance, which creates skepticism about their successful developments. To promote trust model implementation and help practitioners in its redaction, this article provides an outline of what a trust model is, why it is essential, and an example of how it is elaborated. The discussed example comes from a case study of a dairy company that implemented blockchain for the traceability of its products. Despite being tailored on a traceability project, the redaction and elements of the trust model, with few adjustments, could be easily readapted for other applications.

1. Introduction

A recent article (Feb 2022) of the Wall Street Journal supports the view that newly enacted regulations could facilitate the adoption of blockchain technology in the supply chain industry [1]. The idea of exploiting blockchain characteristics to trace real-world assets can be traced back to 2012 with the so-called "colored coins," a means to "attach" commodities to the bitcoin cryptocurrency [2]. Given the limitation of Bitcoin, applications for traceability were then built on Ethereum that allowed a more agile development environment. For example, Walmart, exploiting IBM Food Trust technology, was among the first in 2017 to implement blockchain technology to enhance transparency over its products [2–4]. Later, an ever-growing number of companies proposed blockchain-based traceability systems in their business. However, studies from blockchain observatories show that only a small percentage of them were eventually implemented [5]. Despite the hype and the countless proposals, the successful implementation of blockchain technology into the business requires an in-depth understanding of the technology potential and in particular, of its limitations. When used for real-world applications, and in particular for businesses, blockchain technology often requires "oracles" whose role is rarely considered and evaluated [6]. A 2020 research shows that less than 15% of publications discuss the role of oracles in the academic world, while a bibliometric study focused on oracles research displays only 111 papers till 2021 [7,8]. Neglecting the oracle's role is critical since blockchain applications involving real-world data are not trustless. As a matter of fact, integrating blockchain doesn't imply the acquisition of blockchain properties into the business. Even utilizing a decentralized blockchain, the reliance on a centralized oracle would completely nullify the benefits brought by blockchain technology. Therefore, a detailed and transparent document is needed to explain why the blockchain application is reliable, functional, and trustworthy. This document takes the name of "trust model" [9,10].

Unfortunately, apart from oracle providers whitepapers (e.g., Chainlink, Razor) that outline their trust model, often with a game-theoretical approach, blockchain proposals/applications for businesses rarely provide such a document [11,12]. This contributes to spreading doubts about the feasibility and genuineness of proposals and business integration in general. For example, a famous talk by Andreas Antonopoulos called "bananas on the blockchain" ironically discusses the problem of the improper implementation of blockchain technology into the business [13]. The bitcoin educator explains that business proposals, such as "tracking bananas," should not be handled with blockchain if there is no reason to do so and if not handled appropriately.

This document aims to clarify the limits of blockchain technology when implemented in real-world applications, defining the needs and use of oracles. Therefore, the purpose of the trust model is outlined, discussing its characteristics and features. An example will also be provided from an active blockchain traceability project to guide managers in redacting appropriate trust models and explain their use better. The idea is not to provide an example of a "successful" blockchain integration since the project is still in the experimental phase but of a complete trust model from which the potential of the underlying application can be evaluated.

## 2. Bitcoin and Ethereum

The first successful blockchain application was the Bitcoin cryptocurrency. With bitcoin, it is possible to store, trade, and mint new coins in a trustless, secure, and decentralized way [14]. Despite its potential, the bitcoin blockchain had significant limitations. Bitcoin was, in fact, labeled by its creator as a "peer-to-peer electronic cash system" because its core application was specifically made to fulfill this function [15]. In technical terms, the bitcoin blockchain is called "Turing Incomplete," which means that there are limitations to the operations executable by the machine. Those constraints were deliberately implemented to prevent unwanted or harmful processes from being executed. Due to these constraints, however, it was soon clear that despite its potential (apart from a few scripts), the bitcoin blockchain was not suitable for much more than the management of the underlying cryptocurrency [16].

With the aim of expanding the functionalities of blockchain, a young programmer, Vitalik Buterin (19yo by that time), proposed a new blockchain ecosystem named Ethereum, "The world Computer." Unlike Bitcoin, Ethereum was a "Turing complete" machine; therefore, it "virtually" allowed any application to be deployed [9]. Despite the innovation brought by Ethereum, however, there was still a constraint that prevented the successful exploitation of the technology at a broader scale; the link between the blockchain and the real world [17].

*2.1 The role of Oracles.*

Blockchains are closed ecosystems, and this characteristic is necessary to ensure their security. Although they are said to be "open" in the sense that their content is freely accessible (readable), they cannot be altered (rewritten) with the same degree of freedom. Closed means also that they are entirely isolated from the external world and from other blockchains. Due to this condition of isolation, blockchains are utterly unaware of events happening outside their ecosystems and are natively not provided with means to fetch extrinsic data. Arguably, the inability to gather data from the external world dramatically limits the range of applications that blockchains can execute [18].

Attempting a workaround to this problem, a new actor is introduced within the blockchain ecosystem. Its role is to gather real-world data and feed the smart contract for it to be successfully executed [19]. Being able to connect two separate worlds exactly like Athen's oracle, this actor took the name of "Oracle" [20].

Data gathered with the aid of oracles includes (but is not limited to) the following:

- Lottery winners;

- Price and exchange rate of real/crypto-assets;
- Dynamic data (e.g., time measurements);
- Weather conditions;
- Political events;
- Sporting events;
- Geolocation and traceability information;
- Events in other blockchains.

An example of a basic operation that can be performed with smart contracts is the so-called "atomic swap," which is a trade between two different cryptocurrencies. It is called atomic because it has the characteristic of atomicity for which or is entirely executed or the entire transaction is reverted. In practical terms, it means that it cannot be "partially" executed. While the blockchain is necessary to guarantee the atomicity of the contract, it is not sufficient to perform the operation alone. A swap between two different cryptos requires their exchange rate, which is a kind of data that is not natively available on the blockchain. This data is then provided by an oracle that queries one or multiple trusted sources and feeds the smart contract for it to be successfully executed [21].

In its most basic form, an oracle ecosystem is composed of three parts. The (1) **Data Source** which can be a web API, Sensor, Database or a human aware of a specific knowledge of event. It is the trusted source that provides the data for the smart contract. Only the data collected by the trusted source is used for the smart contract, but not all the data provided is finally exploited. The (2) **Communication Channel** has the aim of securely transferring the data from the data source to the smart contract. It can be a node, or a trusted execution environment, depending on the oracle architecture and purpose. Finally, the (3) **Smart Contract** determines how to digest the external data. It can be digested as it is or with prior computation (e.g., mean). Usually, It also contains the quality criteria for data to be accepted or discarded [8].

Depending on the specific purpose of the blockchain applications, the oracle ecosystem may slightly change [22]. Lately, oracles with multiple data sources and communication channels are preferred. In case of malfunction or data unavailability, they better guarantee the continuity of service [23]. Furthermore, the use of multiple nodes/data sources can help reduce the trust required by the oracle ecosystem itself [18].

*2.2 What is the "oracle problem"?*

Blockchain is hyped since it is said to run transactions in a secure, trustless, and decentralized way. However, while this is true (to a certain extent) for applications such as bitcoin, this is not necessarily the same for all blockchain applications.

The blockchain consensus mechanism is responsible for the trustless data on the blockchain. Transactions should, in fact, be approved by a strict consensus (e.g., Proof-of-Work, Proof-of-Stake) that undisputedly confirms their reliability. Oracles also provide data to the blockchain; however, being their work essential for the smart contract execution, they have the "privilege" to bypass the consensus mechanism and provide data without a global acceptance [24]. This privilege makes oracles able to insert arbitrary data on the blockchain. For that reason, it is crucial that oracles are trustless or at least trusted.

Unfortunately, there is still no accepted method to provide trustless data with oracles; therefore, oracles themselves must be trusted. However, if oracles are trusted third parties, they produce a "counterparty risk" because if unverified data is fed to the oracle or if it is compromised, then the smart contract will work on data that is manipulated and probably untrue [9]. An application using oracles may then fail in two ways. Either if the oracle is trustworthy and well-programmed, but the data is tampered with or wrong at the source. Or, if data is trusted, the system can fail to transmit data due to a malfunction, deliberate tampering, or collusion for selfish purposes. This conundrum, known as the "Oracle Problem," mines the successful development of decentralized applications and represents a real threat for managers implementing blockchain in their business. In sum, it is arguable that real-world blockchain applications are "not trustless" due to the oracle problem. Therefore, a "trust model" is needed to demonstrate the trustworthiness of the implemented Oracle ecosystem [22].

*2.3 What is a trust model?*

Several documents mention the "trust model" construct, explaining why it is crucial [25–27]. However, a proper definition has yet to be formalized. AL-Breiki [10], in accordance with Eskandari et al. [23] argues that sometimes the trust model coincides with the oracle provider whitepaper. However, since those documents are highly heterogeneous, this explanation does not entirely clarify the concept. Naming it "Model of Trust" instead, Lizcano et al. [28] displays it as a scheme that shows how data is collected, verified, and uploaded on-chain (figure 1).

**Figure 1.** Lizcano et al. [28] "Model of Trust" for digital certificates

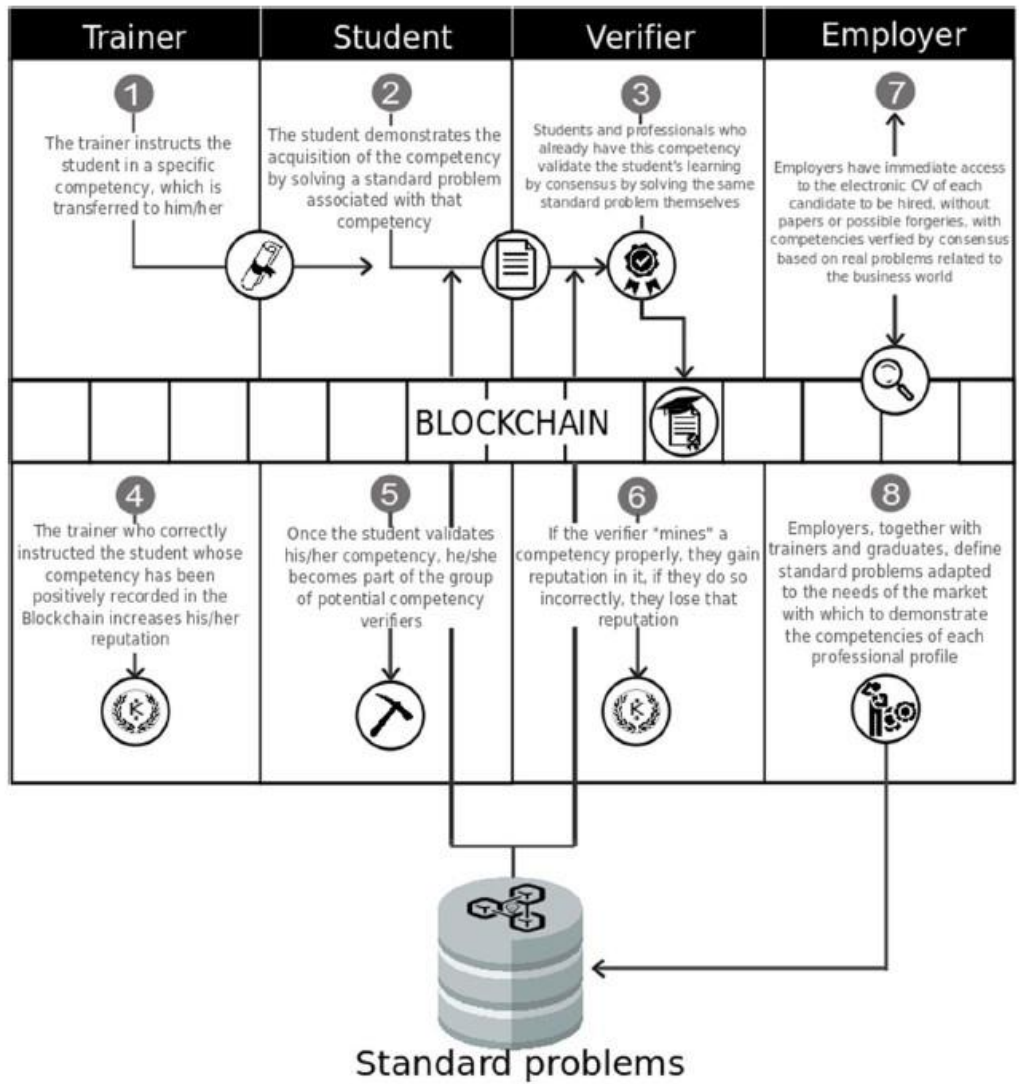

While a comprehensive definition is still not available, understanding the purpose of a trust model is essential and more straightforward.

Since the data collected from oracles is not trustless, the trust model should serve to demonstrate that **data is reliable** in the sense that the selected data source is appropriate for the purpose of the smart contract. **Data is transmitted through a secure channel**, in the sense that it cannot be altered from the moment it is extracted from the data source to the moment it is pushed into the smart contract. Finally that the, **Oracle (or its owner) has no incentive to cheat.** It means that the oracle ecosystem or the entity that manages it cannot (or will not) alter the procedure even if it has the power to. Furthermore, they have no incentive to provide false information for their own benefit.

It is then essential for it to contain at least four elements.

1) **The data validation**: with information about how data is collected and its reliability verified.
2) **The data transmission**: should describe how the data collected is uploaded to the blockchain. Possibly specifying all the passages, with involved actors and used software.
3) **The incentive mechanism**: should specify the relationship within the main involved actors as well as their power to alter the system and deterrents to exercise such a power.
4) **Limitations**: As a trustless oracle ecosystem still does not exist, this section should honestly and transparently describe the conditions under which the system may fail.

In sum, a trust model is a document or a scheme that explains how data is collected, transmitted, and exploited by the blockchain application robustly and transparently. Furthermore, it should formalize the "equilibrium" that prevents the participants from deviating from an honest behavior. Since the process varies almost for any blockchain application, the trust model must be tailored to the specific application even within the same sector. On the other hand, it is arguable that an indefinite number of trust models can be applied and result to be equally robust for the same blockchain application. Limitations should also be considered when selecting the appropriate trust model. A company should, in fact, select the model whose limits are less likely to trigger a negative effect on their blockchain application.

There is no "one-size-fits-for-all" solution; furthermore, there is still not an accepted standard for trust models. Therefore, what is perceived as a solution for someone, may not be an objective solution for everyone else.

The business model canvas can be taken as a similar example to understand a trust model and its purpose better. As explained in the famous book by Osterwalder and Pigneur [29], the canvas can be used to outline and communicate the value creation model to the stakeholders for them to evaluate investment and cooperation opportunities better [30]. Similarly, a trust model can be thought of as a tool to show why the blockchain application should be trusted. Therefore, when a blockchain-based project is presented to potential users and investors, the trust model will provide a broader range of information that may help legitimate projects to obtain more visibility and frauds to be recognized more easily. A study from the Pennsylvania University about the fifty main Initial coin offerings (ICOs) by the amount of raised capital showed that a considerable portion of the projects was not even programmed for the intended purpose [31]. The request for a further document such as a trust model may effectively address this issue. Indeed, there can be the case that the presented trust model is eventually not implemented in the project. However, the further effort required to create and outline a trust model should constitute a deterrent for fraudulent projects to be pursued. The proof-of-work mechanism, for example, does not guarantee honest behavior by censoring inappropriate practices but by requesting an effort that makes dishonest behavior mostly inconvenient [32].

## 3. Formalization of a trust model: A case study

To help practitioners in its redaction, this article also discusses an example of a trust model made for a working blockchain-based application. The application concerned the traceability of dairy products for economically sustainable purposes and was supervised by the University of Verona department of business administration. The following information provides some context to the case study.

*3.1 The case study*

The blockchain traceability project started in 2018 from the cooperation of a dairy company in northern Italy and the University of Verona. The dairy company was founded in 1966 by a congregation of breeders with the aim of producing homogeneous local food. Trusting the quality of their products, they competed at a national and international level to raise awareness about their product's excellence. Since then, the cooperative has received countless prizes, including the Caseus Veneti and the World Cheese Award. Lately, it has also been included in the Super Gold Ranking of "Asiago DOP (Fresh & Aged)," and is thus listed among the best cheeses in the world. As the company realized that its products were being counterfeited, it decided to use blockchain to add proof of authenticity (PoA) for its clients. It is nearly impossible for authorities to spot counterfeit goods outside their jurisdiction domain, then the PoA would at least help customers recognize them. The company relied on an external IT consultant to upload data about dairy production on the blockchain and put a QR code on its product so they could be scanned for provenance and authenticity verification.

*3.2 Oracle problem characteristics in supply chain applications.*

The conceptualization of a trust model serves to overcome the impact of the oracle problem in the specific blockchain application. Therefore, it is necessary to outline first what are the limitations determined by the use of oracles in the supply chain in general and then in the case under analysis.

The use of blockchain in the supply chain has been proposed because since with this technology, it is possible to trace the provenance and every movement of a cryptocurrency with a high level of reliability, it was hypothesized that its application on tangible assets would have led to a similar level of reliability. Unfortunately, since blockchains are closed ecosystems, a real-world asset cannot be permanently attached to the blockchain, nor a direct link can be established. Therefore, data regarding real-world assets should be transmitted to the blockchain using oracles. The use of this workaround leads to the following outcomes:

1) The same level of traceability of cryptocurrencies is unlikely to be replicated with the use of oracles. It would mean that there should be an oracle registering and uploading data on the blockchain for every movement of tangible goods. On the other hand, even hypothesizing the availability of all these oracles, the transaction costs for registering all this data on the blockchain in traditional ecosystems (e.g., Ethereum) would hardly be profitable.
2) Since the data about the products is under the control of the producing company, and oracles are managed or contracted by the producing (or distributing) company, there is no reason to hypothesize on a priori belief that the data uploaded on the blockchain is trustworthy and reliable. Suppose there is a problem with the provenance of a product that would damage the company's image. In that case, it is improbable that this data will be voluntarily uploaded on the blockchain.
3) While digital assets can be hashed and the hash registered on the blockchain or directly minted as Non-Fungible Tokens, tangible assets cannot exploit these opportunities. The attachment of a real-world asset to the blockchain is still debated since there is no secure and stable way to link those two worlds physically. Unfortunately, common methods such as QR codes, NFT tags, and holograms can still be manipulated.

Therefore what is possible to do with the blockchain in the supply chain sector with the existing technology is register product data on the ledger and have it stored immutably and transparently so that it is publicly accessible. Of course, the process of data gathering, transmission, and exploitation have to be always formalized, with a dedicated trust model.

*3.3 Trust model conceptualization*

As stated in paragraph 2, the trust model should outline: (1) how data is gathered and why it is reliable, (2) how it is securely transmitted to the blockchain, (3) why oracles have no incentive to cheat (4), what are the limitations of the proposed approach.

Concerning the case study, the discussion over these requirements was elaborated as follows:

DATA VALIDATION: The company has the data about the products under its control. Theoretically, it is in the best position to decide what information to write on the blockchain and, therefore, to manipulate product data provenance. However, certain types of quality products are subject to the supervision of a third-party authority that provides provenance certifications (e.g., D.O.P., D.O.C.G.). Therefore, the idea is to register on the blockchain only products whose provenance is certified by a third party on which the company has no authority. In this case, the certification authority (D.O.P.) has a long history of product certifications for its high-quality standards, and it is well-known worldwide. Therefore the reliability of product data is ensured by the certification authority and not by the producing company. The idea is to redirect the trust to an actor whose credibility is undisputed. In this case and only for the chosen product, it is the D.O.P. certification authority. Choosing a poor or unrelated certification authority or creating one for the specific purpose of certifying blockchain traced products would not grant the same level of credibility.

DATA TRANSMISSION: The company does not autonomously transmit the data about products on the blockchain but relies on a third-party IT consultant specialized in providing support for blockchain applications. The role of the consultant is essential to compensate for the limited knowledge of the dairy company in blockchain technology but also to perform a double check on the provided product data. The dairy company, in fact, provides all the data considered worthy of being uploaded on the chain. However, the consultant selects the information that is actually required for the product provenance and, if necessary, asks the dairy company for further data. The whole process required a reorganization of data process and storage as well as an integration with the CRM software for data to be directly available to the consultant. In this specific case, the data upload is entirely entrusted to the consultant, although the company may perform further checks after the data upload.

INCENTIVE MECHANISM: Either the company, the consultant or the certification authority may deliberately contribute to providing erroneous data on the blockchain. However, the chance for these events to happen is remote due to the following reasons:

1) The certification authority was not explicitly created to certify products on the blockchain and has no partnership with the dairy company and the consultant. The role of the authority is to supervise producers and certify that their products match or pass the desired standards. An agreement between the supervisor and the company to falsify the product information, although possible, would undermine the credibility of the certification authority, which would lose the trust of other companies and worldwide customers. On the other hand, mistakes in supervising data should be taken into consideration.
2) Despite the fact that the company has its product data supervised by a third party, it can still decide to upload different data on the blockchain since the certification authority does not handle this passage. If that happens, the data uploaded on the blockchain would be immutably stored on the ledger and freely accessible by anyone for auditing purposes. Therefore, in the case of dispute, the company would be unable to deny the manipulation of data. Nonetheless, although it is unlikely to provide erroneous data voluntarily, the chance of a mistake still exists.
3) Lastly, the consultant company having the role of transmitting the data to the blockchain could manipulate product information before the final upload. Again, although virtually possible, this eventuality is remote as it would mean losing the consultant contract with the dairy company and the credibility as a consultant, probably jeopardizing any future collaboration with other companies. Although there is no plausible reason for a deliberate data manipulation from the consultant, a software/hardware failure may still alter the provided data. However, this should represent an even less likely scenario for IT, specialized companies.

A scheme that summarizes the defined elements of the conceptualized trust model can be retrieved in figure 2.

**Figure 2.** Application Trust Model

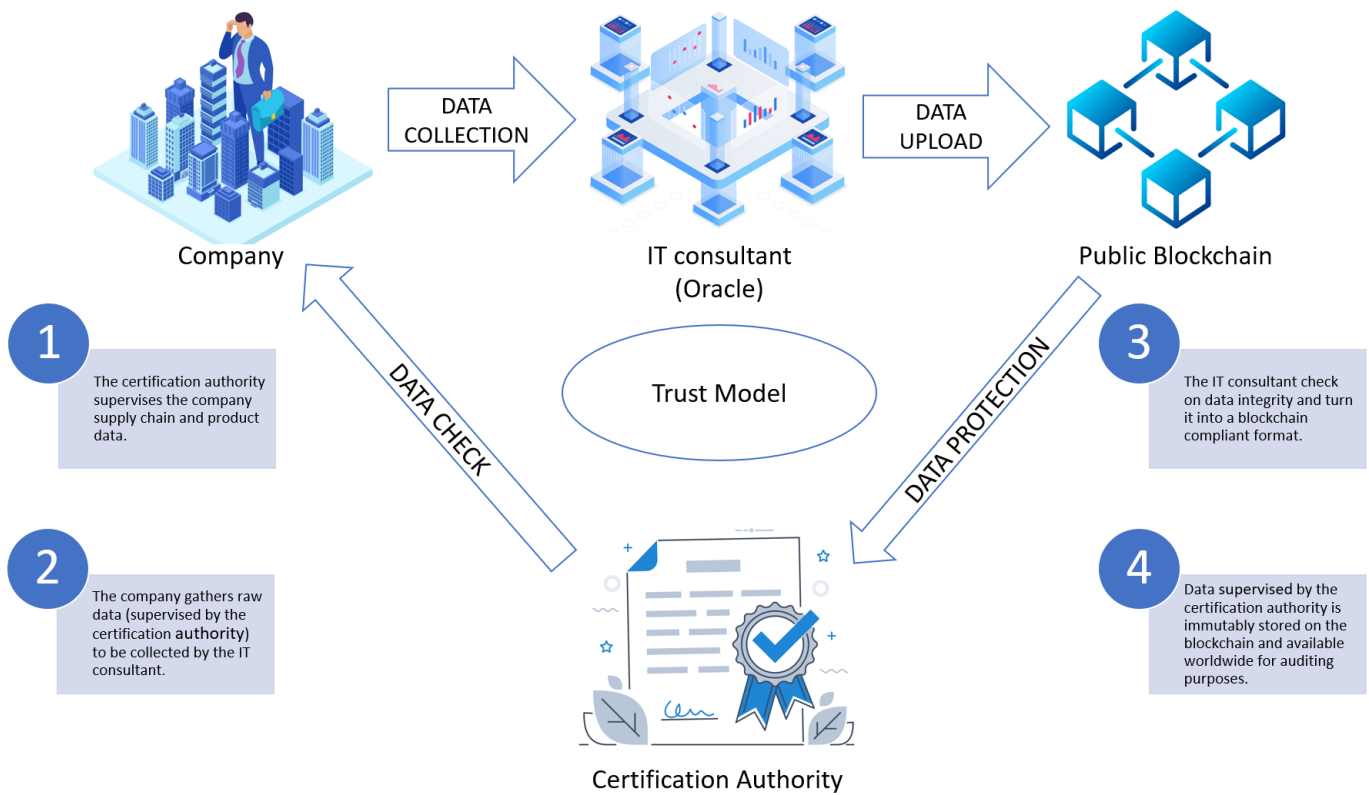

The above graphical representation of the trust model, similarly to the one presented by Lizcano et al. [28], also shows the purpose of the blockchain application, but it is not always the case. As shown, the application objective is to expand the border of the data protection, placing a QR code on the product package that identifies genuine products. Once scanned, the client would get authenticity feedback, and at the same time, the company would obtain the client's data and product location. Being hard to detect fake products outside the national borders, this method should provide an additional verification/protection mechanism.

*3.4 Application and trust model limitations.*

The initial idea of the dairy company was to propose a product traceability system entirely verifiable on the blockchain. However, a dedicated case study and similar researches show that this outcome is not reasonably achievable with the available technology [33,34]. A more realistic result is an on-chain "proof of provenance," with product data verified by a trusted authority. Further product tracing information can be made available with the aid of IoT and cloud computing for data to be registered off-chain. This would allow a dynamic product and customer data management at probably lower registration costs. Furthermore, given the immutability of the ledger, customer information could not be stored on-chain due to GDPR requirements. The management of personal data with blockchain has, in fact, been highly debated in sectors such as E-government and Education [35,36]. To comply with the GDPR and further privacy requirements, sensitive data is usually stored on an off-chain database for it to be permanently deleted upon user request [37,38].

Regarding the link between the blockchain and the physical product, the company opted for a QR code on the external package of the cheese. Initially, there was the idea to print the QR directly on the cheese peel, but the code would have probably been damaged during the cutting and packaging phase making it illegible. Furthermore, when sold in packages, not all the pieces would have been labeled with the QR code. Therefore, the company decided to put the QR code directly on product packages. However, while that choice guarantees the presence of the QR code on all pieces, it does not prevent it from being counterfeited. As specified, it can still be cloned and affixed on a non-genuine product package. Scanning a genuine QR code affixed on a counterfeited product will then erroneously confirm the authenticity of a product. In this specific case study, however, this limitation can be partially overcome due to the fact that fresh products have an early expiration date. Counterfeiting wrong or old codes will display expired products making fake products easier to spot.

As with any equilibrium, the equilibrium found among the actor's choices assumes rationality. While the one presented admits the chances of mistakes, it does not consider the opportunity of irrational behavior. To be realistic, however, the hypothesis of irrational behavior is objectively remote, while it exists, on the other hand, the chance of human mistake. Given the complexity of the operation, a mistake could be expected (although still improbable), by the dairy company in selecting the appropriate data to be sent to the IT consultant. Aware of that eventuality, the dairy company is investing in automating the data collection process, also with the help of specialized consultants.

Finally, a discussion on the blockchain application outcome is required. It must be said that quite a similar result could have also been achieved with other technologies and without involving blockchain. Providers such as Certilogo successfully enable product authentication utilizing Artificial Intelligence to retrieve product provenance data [39].

However, integrating blockchain in the process makes it possible to obtain two further advantages. First, the immutability of data guarantees that information about product provenance is not altered once registered on the ledger. It means that even if the company disappears, it will always be possible to perform an audit on traceability and authenticity data. Second, in a perspective vision, the blockchain integration will grant features such as company tokens and NFT, which should eventually support the metaverse product versions.

4. Conclusion

This article provides a description of the trust model, as well as its needs and purpose. It aims to serve as a guide for managers to help elaborate, redact and present a trust model for their blockchain-based application. In order to clarify its importance, the features of bitcoin and Ethereum are discussed as well as the limitations of smart contracts and the use of oracles. The idea of the trust model is then outlined, explaining what it is, why it is essential, and how it is conceptualized. Therefore, an example of a trust model for a traceability application is also provided, discussing its elements as well as its features and limitations.

The idea is that a blockchain-based project with a robust, transparent, and well-written trust model should be more reliable for investors as well as for users with respect to projects that neglect this essential component.

This contribution is not meant to be an exhaustive guide to the trust model redaction but as an available resource to build upon. Of course, trust models are still at their infancy stage, and it is possible (and welcomed) that other authors or practitioners elaborate a better "tool" or build on this one proposing improvements. Regardless of how trust models will finally be handled and elaborated, it is sure that as long as the blockchain oracle problem is not solved, there will always be the need for efficient trust models.